\documentclass[a4paper,11pt]{article}
\pdfoutput=1
\usepackage{jheppub}
\usepackage[T1]{fontenc}

\title{Uniform quenching processes in a holographic s+p model with reentrance} \vskip 2cm \vskip 2cm
\author[a]{Chuan-Yin Xia,}
\author[a]{Zhang-Yu Nie,$^1$\note{Corresponding author.}}
\author[b]{Hua-Bi Zeng,}
\author[a]{Yu Zhang}
\affiliation[a]{Kunming University of Science and Technology,\\Kunming 650500, China}
\affiliation[b]{Center for Gravitation and Cosmology, College of Physical Science and Technology, Yangzhou University,\\ Yangzhou 225009, China}
\emailAdd{niezy@kust.edu.cn}
\abstract{We study the homogenous quenching processes in a holographic s+p model with reentrant phase transitions. We first realize the reentrant phase transition in the holographic model in probe limit and draw the phase diagram. Next, we compare the time evolution of the two condensates in two groups of numerical quenching experiments across the reentrant region, with different quenching speed as well as different width of the reentrant region, respectively. We also study the dynamical competition between the two orders in quenching processes from the normal phase to the superconductor phase.}
\begin{document}
\maketitle
\flushbottom

\section{\bf Introduction }
The AdS/CFT correspondence~\cite{Maldacena:1997re,Gubser:1998bc,Witten:1998qj} is a strong-weak duality between a quantum filed system without gravity and a classical gravity system with one more dimension. It is a precise equivalence between the two systems, thus can be applied to study not only static solutions, but also non-equilibrium processes. Varies non-equilibrium progress including dynamical phase transitions~\cite{Bhaseen:2012gg,GKLTZ} and quantum turbulence~\cite{Liu,ACL,CGL,DNTZ,LTZ} are realized in the holographic superfluid model~\cite{Gubser:2008px,Hartnoll:2008vx}, which shed light on the related study in strongly coupled condensed matters.

It is necessary to include multiple order parameters to describe the competition and coexistence in systems with complicated phase structure. These systems could also be modeled in holography with more than one charged fields. In Ref.~\cite{Basu:2010fa} a holographic model with two s-wave orders is studied in the probe limit and an s+s coexisting phase was discovered. This model was further explored in Ref.~\cite{Cai:2013wma} with considering the full back-reaction on metric. Competition and coexistence between two different types of orders are also studied in holographic models, such as the s+p model~\cite{Nie:2013sda,Nie:2014qma} and the s+d model~\cite{Nishida:2014lta,Li:2014wca}. See Ref.~\cite{Cai:2015cya} for a very nice review.

The non-equilibrium process in such a holographic system will also evolve with multiple orders, and show the dynamical competition between different orders, which can not be revealed from only static solutions. On the other hand, to explore laws in the non equilibrium processes, it is helpful to consider some special phase transitions, 
which can be easily realized in holographic model with multiple orders in probe limit. Therefore, we can avoid 
the complicated numerical relativity problems in study of dynamical processes. Recently, a dynamical process of phase separation is realized a holographic model with two s-wave orders in Ref.~\cite{Yang:2019ibe}. Later, the domain wall physics and bubble dynamics are studied in a holographic model with a first order phase transition between two s-wave phases~\cite{Li:2020ayr}. These studies are all based on holographic models with multiple orders in probe limit.

In holographic models with multiple orders, interesting phenomenon such as reentrance can be realized~\cite{Nie:2014qma,Li:2017wbi}. In such a system, one of the order parameter show non-monotonic behavior and only be non-zero in a middle region. Therefore, it would be interesting to study a non-equilibrium quenching process starting from one side of the reentrant region and end at the other side. Since the width of the reentrant region can be tuned, different dynamical processes with the same initial and final states can be compared in this setup, which show special effects of the middle region in dynamical processes.

However, the reentrant phase transition in Ref.~\cite{Nie:2014qma} is realized with considering back reaction on the metric, which is complicated to study dynamical processes. While in Ref.~\cite{Li:2017wbi}, the reentrant phase transition is realized in a region with the value of Gauss-Bonnet parameter beyond the causality constraint. To study the dynamical processes of the reentrant phase transition in a more convenient setup, in this paper, we first realize the reentrant phase transition in the s+p model in probe limit. Then we design quenching processes across the reentrant region to study the influence of the reentrant order on the non-equilibrium process with the same initial and final states.

The rest of this paper is organized as follows. In Sec.~\ref{sect:setup}, we give the setup of the s+p model and study the static solutions. We draw a phase diagram involving reentrant phase transitions which will be studied in quenching processes.
In Sec.~\ref{sect:dynamical}, we give necessary details for studying the time dependent processes in this holographic model and show time dependent value of s-wave and p-wave condensates in dynamical quenching processes.
Finally, We give discussions and conclusions in Sec.~\ref{sect:conclusion}.

\section{Holographic s+p model in probe limit and the static solutions}
\label{sect:setup}
\subsection{The setup of the holographic model}
We consider a holographic model with a complex scalar and a complex vector charged under the same U(1) gauge field in the gravity side, the duel field theory therefore contains one scalar order and one vector order. The full action is
\begin{eqnarray}
S&=&S_G+S_M~, \label{SAll}
\\
S_G&=&\frac{1}{2 k_g^2}\int d^4 x \sqrt{-g} (R-2\Lambda)~, \label{Sgravity}
\\
S_M&=&\frac{1}{e_s^2}\int d^4 x \sqrt{-g} (-\frac{1}{4}F_{\mu\nu}F^{\mu\nu}-D_\mu\Psi^*D^\mu\Psi-m_s^2\Psi^*\Psi
-\frac{1}{2}\rho^\dag_{\mu\nu}\rho^{\mu\nu}-m_p^2\rho^\dag_\mu\rho^\mu)~, \label{Smatter}
\end{eqnarray}
where $F_{\mu\nu}$ is the strength of the $U(1)$ gauge field $A_\mu$. $\Psi$ is complex scalar field and $\rho_\mu$ is complex vector field. $\{q_s,m_s\}$ and $\{q_p,m_p\}$ are the charges and masses of $\Psi$ and $\rho_\mu$, respectively. Both $\Psi$ and $\rho_\mu$ are charged under the U(1) gauge field with
\begin{eqnarray}
D_\mu \Psi=\partial_\mu \Psi - i q_s A_\mu \Psi~,\\
\bar{D}_\mu \rho_\nu=\partial_\mu \rho_\nu-i q_p A_\mu \rho_\nu~.
\end{eqnarray}
$\rho_{\mu\nu}$ is the field strength of $\rho_\mu$ and is given by
\begin{eqnarray}
\rho_{\mu\nu}=\bar{D}_\mu \rho_\nu-\bar{D}_\nu \rho_\mu~.
\end{eqnarray}

This is a simple setup for a holographic system with both s-wave and p-wave orders charged under the same $U(1)$ gauge field. The phase structure of this system with $m_p^2=0$ and $q_s=q_p=1$ can be found in Ref.~\cite{Nie:2014qma}. The reentrant phase transition has been realized there with appropriate value of the back reaction strength. However, investigating the dynamical processes in the system with considering back-reaction on metric is quite complicated. Therefore, we try to realize a reentrant phase transition in probe limit in this work. In the rest of this section, we firstly show details of calculations for static solutions. After that, we introduce how to get a reentrant phase transition by tuning different parameters, and give the $q_p-\rho$ phase diagram which is helpful to understand the different numerical quenching experiments in the next section.

\subsection{Static solutions}
Because we want to study the dynamical quenching processes without doing any numerical relativity, we take the probe limit in this paper. The background geometry can be taken as a $3+1$ dimensional AdS black brane with
\begin{eqnarray}\label{metric0}
ds^2&=&\frac{L^2}{z^2}(-f(z)dt^2+\frac{1}{f(z)}dz^2+dx^2+dy^2)~,\\
f(z)&=&1-(z/z_h)^3.
\end{eqnarray}

This metric is convenient for solving static solutions, and can be easily transformed to the Eddington coordinates which is the better choice for the dynamical processes. In this metric, $z$ is the radial coordinate of the bulk, with $z=0$ the location of the AdS boundary and $z=z_h$ the horizon. $L$ is the AdS radius. The temperature of the black brane is related to $z_h$ as
\begin{equation}\label{TemperatureE}
T=-\frac{f'(z)}{4\pi}\Big|_{z=z_h}=\frac{3}{4\pi z_h}~,
\end{equation}
which is also the temperature of dual field theory.

We set the following ansatz for the matter fields
\begin{eqnarray}
\Psi=\Psi_s(z)*z/L~,~A_t=\phi(z)~,~\rho_x=\Psi_p(z)~,
\end{eqnarray}
with all other field components being turned off. The scalar field is dual to an s-wave order in the boundary field theory and the vector field is dual to a p-wave one, therefore we mark the related functions with subscriptions $_s$ and $_p$ respectively. 
With this ansatz, the equations of motion for matter fields in the AdS black brane background read
\begin{eqnarray}
q_s^2 \phi^2 \Psi_s/f+D_s\Psi_s&=&0, \label{eqphi}\\
q_p^2 \phi^2 \Psi_p/f+D_p\Psi_p&=&0, \label{eqpsi3}\\
(q_s^2 \Psi_s^2+q_p^2 \Psi_p^2) \phi/f-\partial_z^2\phi/2&=&0.\label{eqpsix}
\end{eqnarray}
in which
\begin{eqnarray}
D_s&=&(z f'-2 f-m_s^2 L^2)/z^2+f'\partial_z+f\partial_z^2~,
\\
D_p&=&-m_p^2 L^2/z^2+f'\partial_z+f\partial_z^2~.
\end{eqnarray}

There are three sets of scaling symmetries in equations (\ref{eqphi}),(\ref{eqpsi3}) and (\ref{eqpsix}):
\begin{eqnarray}\label{scaling}
L\rightarrow \lambda^{-1} L,~m_s\rightarrow \lambda m_s,~m_p\rightarrow \lambda m_p~,\\
z\rightarrow \lambda^{-1} z,~z_h\rightarrow \lambda^{-1} z_h,~ \phi\rightarrow \lambda \phi,~ \Psi_s\rightarrow \lambda \Psi_s,~ \Psi_p\rightarrow \lambda \Psi_p~,\\
q_s\rightarrow \lambda^{-1} q_s,~q_p\rightarrow \lambda^{-1} q_p,~ \phi\rightarrow \lambda \phi,~ \Psi_s\rightarrow \lambda \Psi_s,~ \Psi_p\rightarrow \lambda \Psi_p~.
\end{eqnarray}
The first and second scaling symmetries can be used to set $L=1$ and $z_h=1$ respectively. One can recover the two parameters to any other values with these two symmetries after we get the numerical solutions. The last scaling symmetry implies that the value of charge coupling parameter will not influence the qualitative behavior of phase transitions in the models with single order. However, in a system with two charged orders, the ratio $q_p/q_s$ does have non-trivial influence on the phase transitions. We work in the probe limit and do not change the background metric in the dynamical processes, therefore we set $L=z_h=q_s=1$ in the rest of this paper.

To solve these coupled equations, we also need to specify the boundary conditions. The expansions near the horizon $z=z_h$ are
\begin{eqnarray}\label{horizon}
\phi(z)&=& \phi_1(z-z_h)+\mathcal{O}((z-z_h)^2)~,\nonumber \\
\Psi_s (z)&=& \Psi_{s0} + \mathcal{O}(z-z_h)~,\nonumber \\
\Psi_p (z)&=& \Psi_{p0} + \mathcal{O}(z-z_h)~,
\end{eqnarray}
while the expansions near the boundary are
\begin{eqnarray}
\phi(z)|_{z=0}&=&\mu-\rho z~,
\\
\Psi_s(z)|_{z=0}&=&(\Psi_{s-}  z^{\triangle_{s-}}+\Psi_{s+} z^{\triangle_{s+}})/z~,
\\
\Psi_p(z)|_{z=0}&=&\Psi_{p-}  z^{\triangle_{p-}}+\Psi_{p+} z^{\triangle_{p+}}~.
\end{eqnarray}
$\triangle_{s\pm}$ and $\triangle_{p\pm}$ can be calculated with the mass parameters as
\begin{eqnarray}
\triangle_{s\pm}&=&\frac{3\pm\sqrt{9+4 m_s^2}}{2}~,\\
\triangle_{p\pm}&=&\frac{1\pm\sqrt{1+4 m_p^2}}{2}~.
\end{eqnarray}

In this paper, we take the standard quantization, in which $\Psi_{s-}$ and $\Psi_{p-}$ are regarded as the source terms of the boundary operators while $\Psi_{s+}$ and $\Psi_{p+}$ are regarded as the vacuum expectation values, with conformal dimensions $\triangle_{s+}$ and $\triangle_{p+}$ respectively. 
We set the conditions $\Psi_{s-}=\Psi_{p-}=0$ to obtain the solutions of spontaneous U(1) symmetry breaking. $\mu$ is the chemical potential and $\rho$ is the charge density of the boundary field theory.

We can see that 
$\Psi_s$ and $\Psi_p$ do not directly coupled with each other in the equations of motion. Therefore we can turn off one of the two functions and get the single condensate solutions consistently. Besides the s-wave and p-wave single condensate solutions, we need to also find solutions with both two condensates non-zero, which are denominated as the coexisting s+p solutions in this paper.

We plan to quench the charge density $\rho$ of the holographic system. Therefore in this section, we report the condensate behavior of typical reentrant phase transition as well as the phase diagram with the horizontal axis $\rho$, which means we work in canonical ensemble. In order to compare the stability of different solutions, we calculate the Gibbs free energy holographically from the Euclidean on-shell action. In the probe limit, the background geometry is the same for different solutions, thus we only calculate the contribution of Gibbs free energy from the matter action~(\ref{Smatter}) to compare the stability. Substituting the equations of motion into the Euclidean on-shell action, we can get the expression of the matter contribution of Gibbs free energy
\begin{equation}
G_m=\frac{V_2L^2}{T} (\frac{\mu\rho}{2}+ \int_0^{z_h}(\frac{q_s^2 \phi^2\Psi_s^2}{f}+\frac{q_p^2 \phi^2\Psi_p^2}{f})dz)~.
\end{equation}

\subsection{Tuning towards a reentrant phase transition and the 
phase diagram}
With the equations of motion and the corresponding boundary conditions, the Newtonian iterative algorithm is used to solve the boundary value problem and get the different solutions. Since we have set $L=z_h=q_s=1$, there are still three parameters \{$m_s^2,m_p^2,q_p$\} left, which can be used to tune the phase structure. In this subsection, we further explain how to tuning the phase transition to be a reentrant one and show the resulting phase diagram. We also give the condensate behavior as well as the relative value of Gibbs free energy for a typical reentrant phase transition.



According to experiences in the holographic study with multiple orders, the coexisting solution exist when the free energy of the two single order solutions are very close to each other~\cite{Nie:2014qma}, which can be explained in the landscape picture~\cite{Li:2020ayr}. In order to have a reentrant behavior for the coexisting phase, we need to get a special relation for the free energy curves for the two single condensate solutions: one of the single condensate solution should be more stable both at the left and right region, while the other single condensate solution should be more stable in the middle region. Therefore, we tune the three parameters \{$m_s^2,m_p^2,q_p$\} to make the free energy curves for the two single condensate solutions tangent to each other or have two intersection points in the middle.

In order to tuning the two free energy curves to be the above configuration, we need to understand the influence of the different parameters on the free energy curves. As shown in Ref.~\cite{Li:2017wbi}, when we tune one of two parameters $m_s^2$ and $q_s$ with other one fixed, the free energy curve of the single condensate s-wave solution will be changed ``parallel''. The similar law holds for the p-wave solution when we tuning the parameter $m_p^2$ or $q_p$.

Our strategy of searching for the reentrant phase transition is similar to that in Ref.~\cite{Li:2017wbi}.
At first, the parameter $q_p$ is always set to the value that the p-wave and s-wave solutions have the same critical value $\rho_c$. Then we can tuning the two parameters $m_s^2$ and $m_p^2$ to get the second intersection point (besides the critical point) of the two free energy curves for the s-wave and p-wave solutions. Based on the above result, we further tuning the value of $q_p$ to get the appropriate reentrant phase transition. It should be noticed that with only the mass parameters and charge coupling parameters, it is not likely to get the reentrant phase transition with two s-wave orders. This is the reason of that we choose one s-wave order and one p-wave order to build the model.

\begin{figure}[tbp]
\includegraphics[width=0.8\textwidth,origin=c,trim=0 260 120 280,clip]{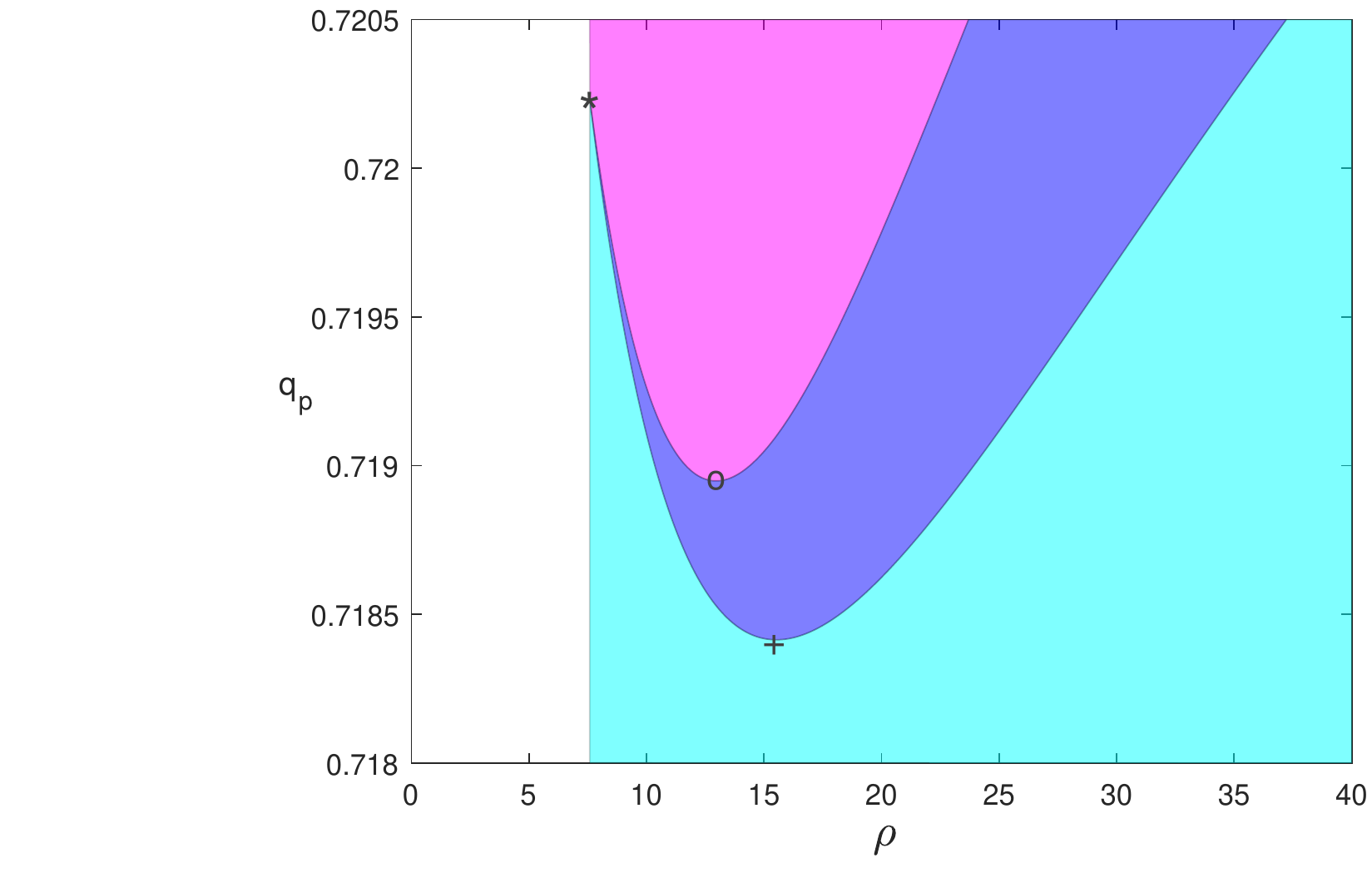}
\caption{\label{phasediagram}  The $q_p-\rho$ phase diagram with $m_s^2=0$ and $m_p^2=3/4$. The horizontal axis is the charge density $\rho$ while the vertical axis is $q_p$. The white region is dominated by the normal phase, the cyan region is dominated by the s-wave phase, the magenta region is dominated by the p-wave phase and the blue region is dominated by the s+p phase. The symbol $*$ marked the quadruple point, while the symbols $o$ and $+$ marked the two minimal points of two curves for the critical points of the s+p phase. The values of $q_p$ for the three points are $q_p^*=0.720234$, $q_p^o=0.718949$, $q_p^+=0.718415$.}
\end{figure}

In order to simplify the numerical work in quenching processes, we prefer the value $\Delta_+-\Delta_-$ for the scalar and vector orders to be integers. Therefore we only consider several discrete values for $m_s^2$ and $m_p^2$. 
Finally, we find a nice choice of the two parameters $m_s^2=0$ and $m_p^2=3/4$ and get the reentrant phase transitions. We fix $m_s^2=0$ and $m_p^2=3/4$ in the rest of this paper and show the $q_p-\rho$ phase diagram in Figure.~\ref{phasediagram}.

We can see from Figure.~\ref{phasediagram} that there are four regions dominated by the normal phase, the s-wave phase, the p-wave phase and the s+p phase respectively. The lines separating different phases are all second order critical points. Because we only tune the parameter $q_p$, the critical value $\rho_c$ for the s-wave phase transition is not changed, as a result the line between the white and cyan region is vertical. The line between the white and magenta region is not vertical and show the dependence of $\rho_c$ for the p-wave phase transition on $q_p$.

We can separate the phase diagram to four part by the three typical values of $q_p$. When $q_p<q_p^+$, the s-wave phase always dominate in the large $\rho$ region. When $q_p^+<q_p<q_p^o$, the system undergoes a reentrant phase transition with non-monotonic p-wave condensate. This is the case we search for, and doing quench with in the next section. When $q_p^o<q_p<q_p^*$, the system contains two sections of s+p phases. This can be also understood as reentrance for the non-monotonic behavior of p-wave condensate. When $q_p>q_p^*$, the system contains a typical s+p phase of the `x-type'. For large enough value of $q_p$, the system is also possible to be always dominated by the p-wave phase in the large $\rho$ region, which is not shown in this phase diagram.

In the rest of this section, we set $q_p=7189/10000$ to obtain a typical reentrance phase transition in the region $q_p^+<q_p<q_p^o$. We show the condensate values as well as the relative values of Gibbs free energy with respect to $\rho$, which give a concrete example for the static properties of reentrant phase transitions.

\begin{figure}[tbp]
\includegraphics[width=0.8\textwidth,origin=c,trim=0 260 120 280,clip]{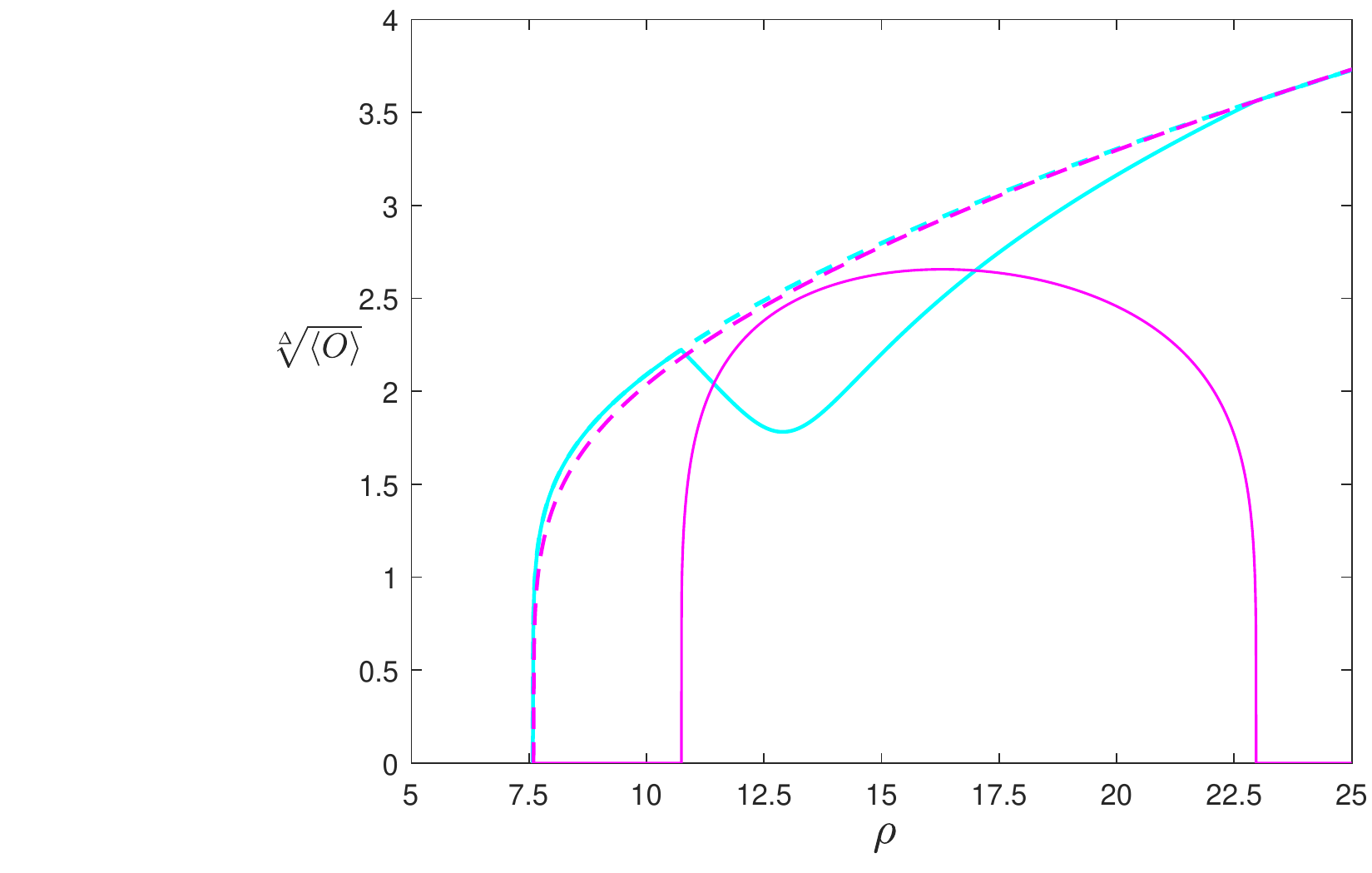}
\caption{\label{condensate}  The condensate values versus the charge density $\rho$ with $q_p=7189/10000$. The solid lines represents the condensate values for the most stable solution with both the s-wave and p-wave orders turned on. The dashed lines represents the condensate values for the solutions with only single condensate. The lines colored cyan is for the s-wave order, and lines colored magenta is for the p-wave order. }
\end{figure}

The condensate values of the two orders for the different solutions are given in Figure.~\ref{condensate}. We can see from this figure that, when we increase the value of $\rho$ from a small value very slowly, the s-wave order condensed firstly. If there is only p-wave order, the p-wave condensate will form at a slightly larger value of $\rho$. In another word, the critical value $\rho_{c-s}$ for the s-wave solution is slightly smaller than the critical value $\rho_{c-p}$ for the p-wave solution. In the s+p model where both the s-wave and p-wave orders are turned on, the s-wave condensate prevent the formation of p-wave condensate until the left critical point for the s+p solution $\rho_{c-sp-L}$. However, the formation of the p-wave condensate is not monotonic with the increasing of $\rho$. The value of the p-wave condensate increase to a maximum value, then decrease and finally vanish at the right critical point $\rho_{c-sp-R}$, which is a typical curve of the order parameter in a reentrant phase transition. Because the curve of the p-wave condensate from a shape of letter `n', we call this kind of phase transition as `n-type'.

\begin{figure}[tbp]
\includegraphics[width=0.8\textwidth,origin=c,trim=0 260 120 280,clip]{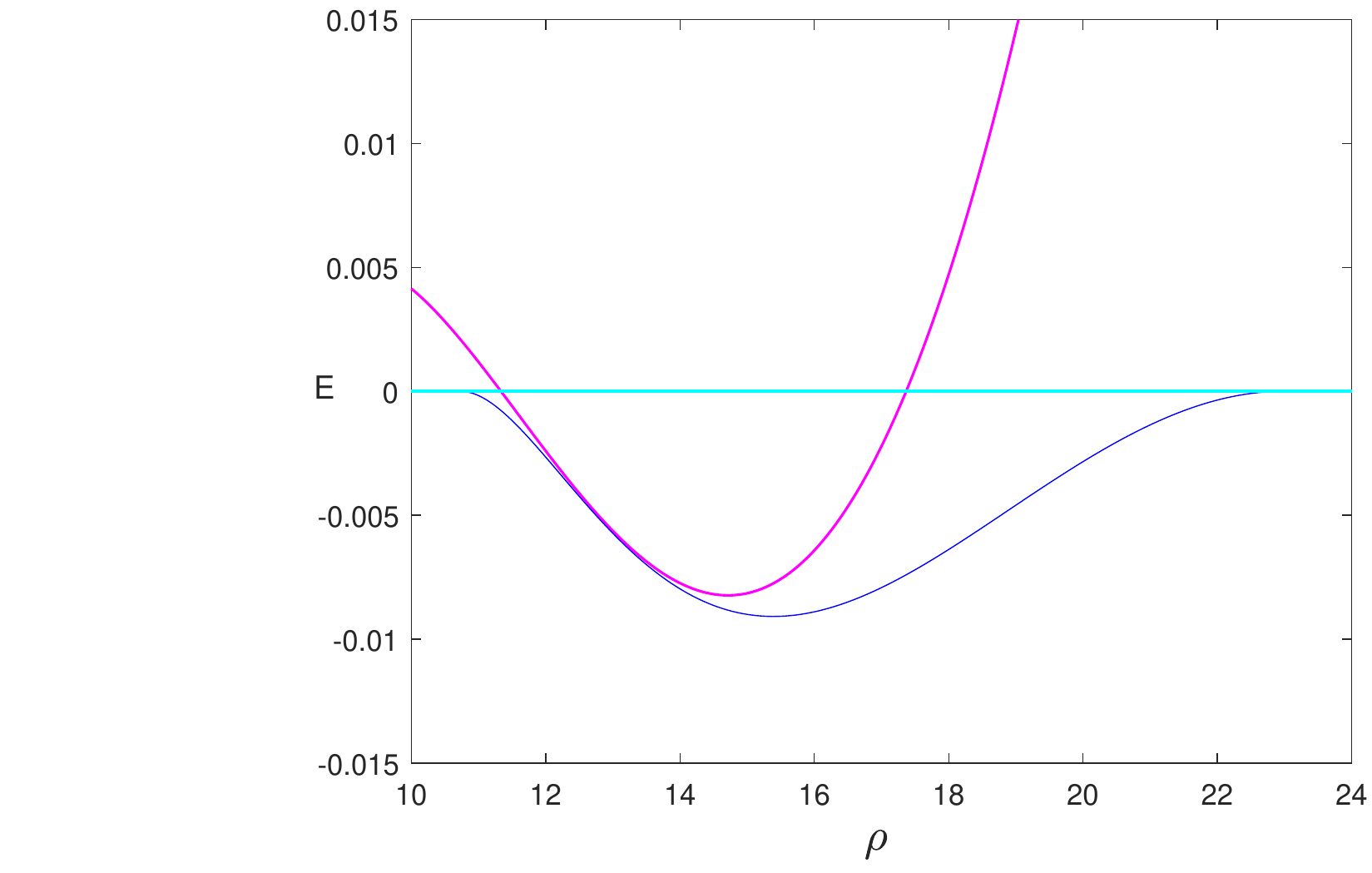}
\caption{\label{FreeE}  The relative value of Gibbs free energy $G-G_s$ with respective to $\rho$. $G_s$ is the Gibbs free energy of the s-wave solution, therefore the cyan line for the s-wave solution coincide with the horizontal axis. The magenta line denotes for the p-wave solution, and the blue line denotes for the coexisting s+p solution.}
\end{figure}

We also draw a figure for relative values of Gibbs free energy for the three different condensed solutions in Figure.~\ref{FreeE}. Because we work in probe limit, only the contribution from the matter action (\ref{Smatter}) differs for the different solutions. The absolute values of this matter contribution for different solutions are very close to each other, therefore we draw the relative values $G-G_s$, where $G_s$ is the Gibb free energy for the s-wave solution. As a result, the line for the s-wave solution in Figure.~\ref{FreeE} is horizontal.

We can see from Figure.~\ref{FreeE} that, the Gibbs free energy of the s-wave solution is lower in both the left(small $\rho$) and right(large $\rho$) region, while the Gibbs free energy for the p-wave solution is lower than that of the s-wave solution in the middle region. The free energy for the s+p solution is the lowest in the region where it exist and the curve is tangent to the curve for the s-wave solution at the two critical points. This is a typical free energy relation for the reentrant phase transitions with $q_p^+<q_p<q_p^o$, and explains the non-monotonic behavior of the condensate values in Figure.~\ref{condensate}.

If we change the value of $q_p$, the Gibbs free energy curve for the s-wave solution will not be affected, while the curve for the p-wave solution will be `parallel' moved upwards or downwards. As a result, the width of the s+p phase will decrease or increase, as shown in Figure.~\ref{phasediagram}. Therefore it is convenient to compare the influence of the width of the reentrant region on the quench process with the same starting and end points.
\section{Dynamical processes}
\label{sect:dynamical}
In this section, we study quenching processes in this holographic s+p model with $m_s^2=0$ and $m_p^2=3/4$. For simplicity, we only consider homogeneous quenching processes with linearly increasing charge density $\rho(t)=\rho_0+v t$, where $t$ is the time coordinate, $\rho_0$ is the initial value of charge density and $v$ is the quench rate.

We first study quenching processes across the reentrant region with $q_p^+<q_p<q_p^o$, starting from the left side of the reentrant region and ending at the right side. The value of $q_p$ controls the width of the reentrant region, while the quench rate $v$ controls the speed of quenching processes. We compare the effects of the two parameters in two groups of numerical experiments, respectively.

At the end of this section, we also quench the holographic system from the normal phase to compare the evolution of the two orders in this s+p model and that in model with single condensate. 
\subsection{Dynamical equations in Eddington coordinates}
In order to quench the holographic system, it's more convenient to transform to the ingoing Eddington coordinates
\begin{equation}
ds^2=\frac{L^2}{z^2}(-f(z)dt^2-2dtdz+dx^2+dy^2))~.
\end{equation}

If we transform the static solution in the previous coordinates with metric (\ref{metric0}) into the one in Eddington coordinates, the component $A_z$ of the U(1) gauge field will become nonzero. However, we can use the gauge symmetry to set again $A_z=0$, at the cost of a complex valued scalar field $\Psi$. Therefore we rewrite the anstatz as
\begin{equation}
\Psi=\Psi_s(z,t)*z/L, A_t=\phi(z,t), \rho_x=\Psi_p(z,t)~,
\end{equation}
where both $\Psi_s(z,t)$ and $\Psi_p(z,t)$ are complex valued. The resulting time dependent equations of motion are
\begin{eqnarray}
i q_s (2\phi\partial_z + \partial_z\phi)\Psi_s + D_s \Psi_s - 2\partial_t\partial_z\Psi_s=0\label{eqs}~,\\
i q_p (2\phi\partial_z + \partial_z\phi)\Psi_p + D_p \Psi_p - 2\partial_t\partial_z\Psi_p=0\label{eqp}~,
\\
\label{eq3}
i q_s(f\Psi_s^\ast\partial_z\Psi_s-\Psi_s^\ast\partial_t\Psi_s - c.c.)-2|q_s\Psi_s| ^2+p.p. -\partial_t\partial_z\phi=0~,
\\
\label{eq4}
q_s(\Psi_s^\ast\partial_z\Psi_s - c.c.) + p.p. - \partial_z^2\phi=0~.
\end{eqnarray}
For simplicity, when the terms of the s-wave part have been written out, we use $p.p.$ to express the terms for the p-wave part, which can be get by simply replacing the index s with p. Similarly, c.c. represents complex conjugation.

The above four equations form a constraint equation~\cite{Zeng:2018ero}
\begin{equation}
\label{eq5}
i q_s(\Psi_s^\ast Eq.(\ref{eqs})-c.c)+p.p=\partial_z Eq.(\ref{eq3})-\partial_t Eq.(\ref{eq4})~,
\end{equation}
which implies that when s-wave and p-wave orders are all turned on, the U(1) current is still conserved.

The boundary expansions are almost identical to the static case, except that all expanding coefficients also depend on time. The expansions near the horizon are
\begin{eqnarray}\label{horizonT}
\phi(z,t)&=& \phi_1(t)*(z-z_h)+\mathcal{O}((z-z_h)^2),\nonumber \\
\Psi_s(z,t)&=& \Psi_{s0}(t) + \mathcal{O}(z-z_h),\nonumber \\
\Psi_p(z,t)&=& \Psi_{p0}(t) + \mathcal{O}(z-z_h)~,
\end{eqnarray}
while the expansions near the boundary are
\begin{eqnarray}
\phi(z,t)|_{z=0}&=&\mu(t)-\rho(t) z
\\
\Psi_s(z,t)|_{z=0}&=&(\Psi_{s-}(t)  z^{\triangle_{s-}}+\Psi_{s+}(t) z^{\triangle_{s+}})/z
\\
\Psi_p(z,t)|_{z=0}&=&\Psi_{p-}(t)  z^{\triangle_{p-}}+\Psi_{p+}(t) z^{\triangle_{p+}}~.
\end{eqnarray}
In the time dependent case, we still set the source free boundary conditions $\Psi_{s-}=\Psi_{p-}=0$, and quench the value of $\rho$ by directly set $\rho(t)$ a time dependent function. The expectation value of the s-wave and p-wave order parameters can still be read from $\Psi_{s+}(t)$ and $\Psi_{p+}(t)$ respectively.

In addition to the boundary conditions, we also need initial conditions. Because we quench the system from a static solution at $\rho=\rho_0$, we set the initial value for the three fields as
\begin{equation}
\phi(z,t)|_{t=0}=\phi_0(z)~,
\end{equation}
\begin{equation}
\Psi_s(z,t)|_{t=0}=\psi_{s0}(z)~,
\end{equation}
\begin{equation}
\Psi_p(z,t)|_{t=0}=\psi_{p0}(z)~,
\end{equation}
where $\phi_0$, $\psi_{s0}$ and $\psi_{p0}$ satisfy the static equations of motion with $\rho=\rho_0$.

We use the fourth-order Runge-Kutta algorithm to solve the time dependent equations of motion numerically. We also add small fluctuations at each step to model thermodynamic fluctuations. 
\subsection{Quenching across the reentrant region with different width}
According to the phase diagram in Figure.~\ref{phasediagram}, the value of $q_p$ controls the width for the reentrant region. In order to focus on the influence of the width of the reentrant region, we fix the positions(values of $\rho$) of the starting point and end point while tuning the value of $q_p$ to get different width of the reentrant region.


The starting point is set at the static solution at $\rho_0=10$, which is dominated by the s-wave phase, thus the initial value of the functions are the s-wave solution with $\Psi_{p0}=0$. The quenching process stop at the time when $\rho(t)$ reaches the final value $\rho(t)=\rho_f=25$, again in the region dominated by the s-wave phase in the static phase diagram. After the end of quenching, the system will go on evolution until it goes to thermal equilibrium on the s-wave solution. 
In such quenching processes, the detailed evolution depend on $q_p$, which control the width of the reentrant region, as well as the quench rate $v$. We first study the influence of the width of the reentrant region by fixing $v$ and set $q_p$ to different values, and later show the effect of $v$ by fixing $q_p$ and set $v$ to different values in the next subsection.

\begin{figure}[tbp]
\includegraphics[width=0.8\textwidth,origin=c,trim=0 260 120 280,clip]{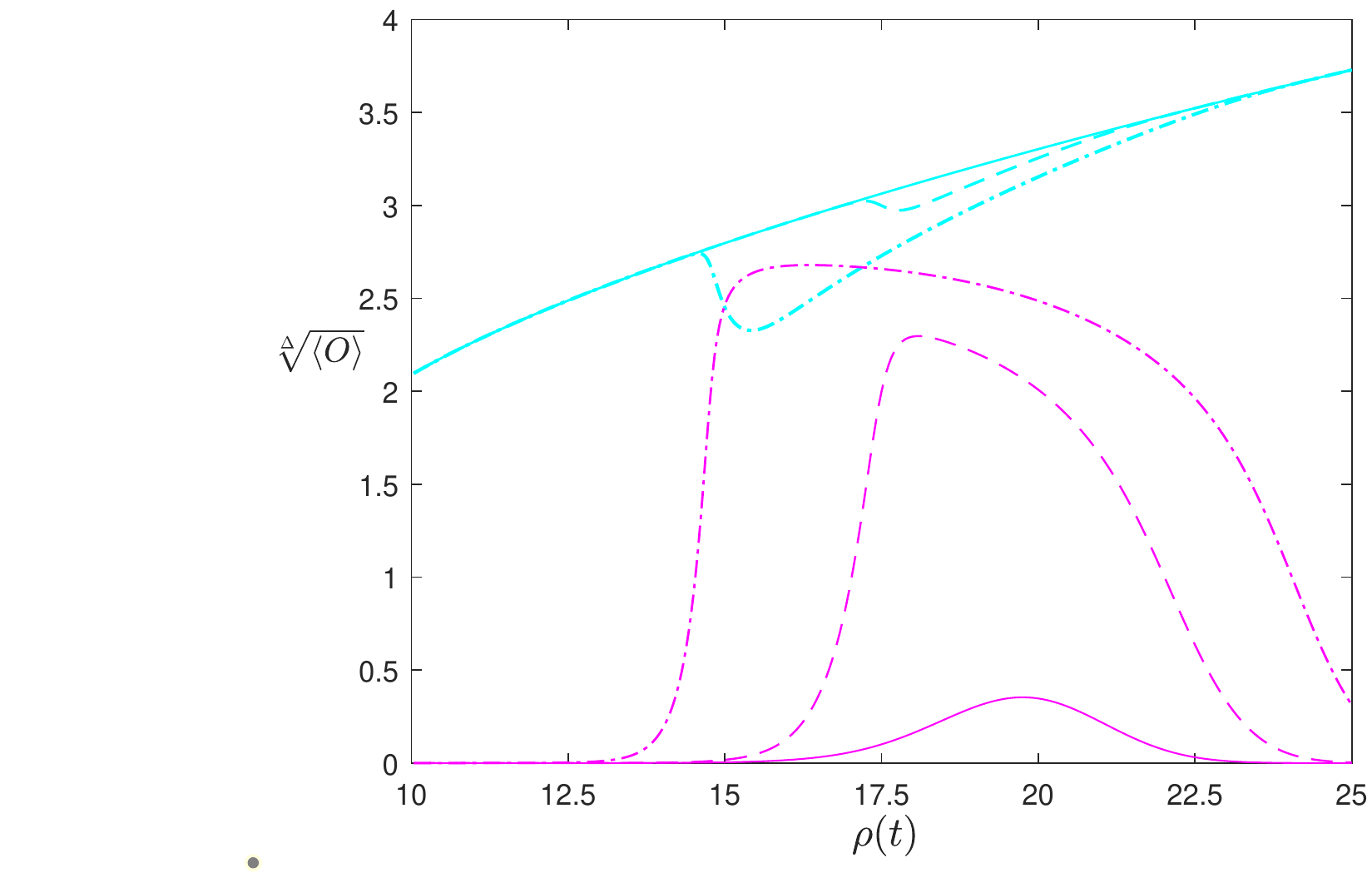}
\caption{\label{quench1} The values of condensates for s-wave and p-wave orders in three quenching processes with $v=0.00005$ and different values of $q_p$. The horizontal axis show the linearly time dependent charge density $\rho(t)$, while the vertical axis show the condensate values $\sqrt[\Delta]{\langle O\rangle}$. Cyan curves denote for the s-wave order and magenta curves denote for the p-wave order. The dotted dashed lines, dashed lines and solid lines denote the three different processes with $q_p=0.7189$, $0.7187$ and $0.7186$ respectively.
}
\end{figure}

We fix the quench rate to $v=0.00005$ and show the time dependent condensate values of the s-wave and p-wave orders in Figure.~\ref{quench1} for three different quenching processes with $q_p=0.7189$, $0.7187$ and $0.7186$ respectively.

In Figure.~\ref{quench1}, we use $\rho(t)$ instead of $t$ as the horizontal axis. 
Cyan curves denote for the s-wave order and magenta curves denote for the p-wave order. The dotted dashed lines, dashed lines and solid lines denote the three different processes with $q_p=0.7189$, $0.7187$ and $0.7186$ respectively. We can see that in all the three processes, the condensate value for the p-wave order increase from zero and decrease after reaching a maximum value, while the s-wave order is suppressed when the p-wave order becomes large. 
In the quenching process with a larger value of $q_p$, the reentrant order get a larger condensate value profile. This is in accordance with the fact that in the static phase diagram the reentrant region is wider with a larger value of $q_p$.

It is obvious that the quenching process with a larger value of $q_p$ get a larger remaining value of p-wave condensate at the end of quenching with $t=t_e$, therefore it seems more time should be taken for the system to go to the final equilibrium state. However, we should notice that although the different quenching process start from the same initial state and end with the same final state in the sense of static solutions, the different values of $q_p$ still make the final states different in the context of non-equilibrium physics. Especially, the quasi-normal modes of the p-wave order depend on $q_p$, therefore the three different quenching processes get different values of relaxing time $\tau$, which can be defined from
\begin{equation}
O-O_f \propto exp(-t/\tau).
\end{equation}
$O$ is the time dependent expectation value and $O_f$ is the expectation value for final equilibrium state. The value of $\tau$ determines the late time relaxing to the final thermal state. The relation between the quasi-normal modes and the late time evolution has already been revealed holographically in the context of dynamical phase transitions in Refs.~\cite{Bhaseen:2012gg,GKLTZ}. In this paper, we numerically fit the late time behavior of the p-wave condensate to get the values of $\tau$.

\begin{figure}[tbp]
\includegraphics[width=0.8\textwidth,origin=c,trim=0 260 120 280,clip]{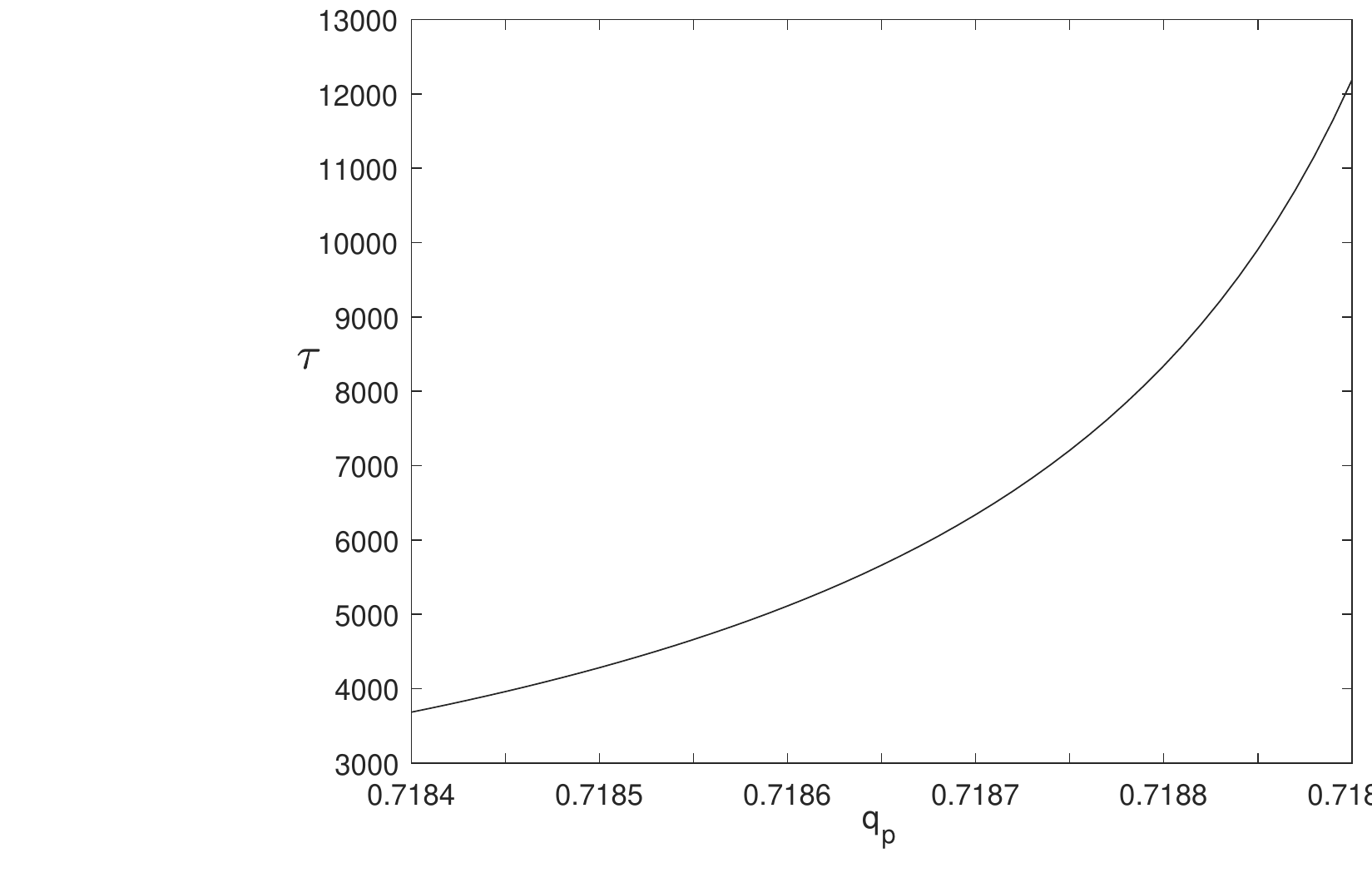}
\caption{\label{tauqp}  The $\tau-q_p$ relation with the final state at $\rho_e=25$.
}
\end{figure}

We plot the curve of $\tau-q_p$ relation in Figure~\ref{tauqp}. We can see that $\tau$ is a monotonic increasing function on $q_p$. Therefore, it takes more time for a perturbed state to go back to equilibrium at $\rho=25$ with a larger value of $q_p$. Together with the previous result of remaining value of p-wave condensate at $t=t_e$, we can conclude that for the quenching processes in this subsection, it takes more time to equilibrate with a larger value of $q_p$ .




\subsection{Quenching across the reentrant region with different quench rate}
\begin{figure}[tbp]
\includegraphics[width=0.8\textwidth,origin=c,trim=0 260 120 280,clip]{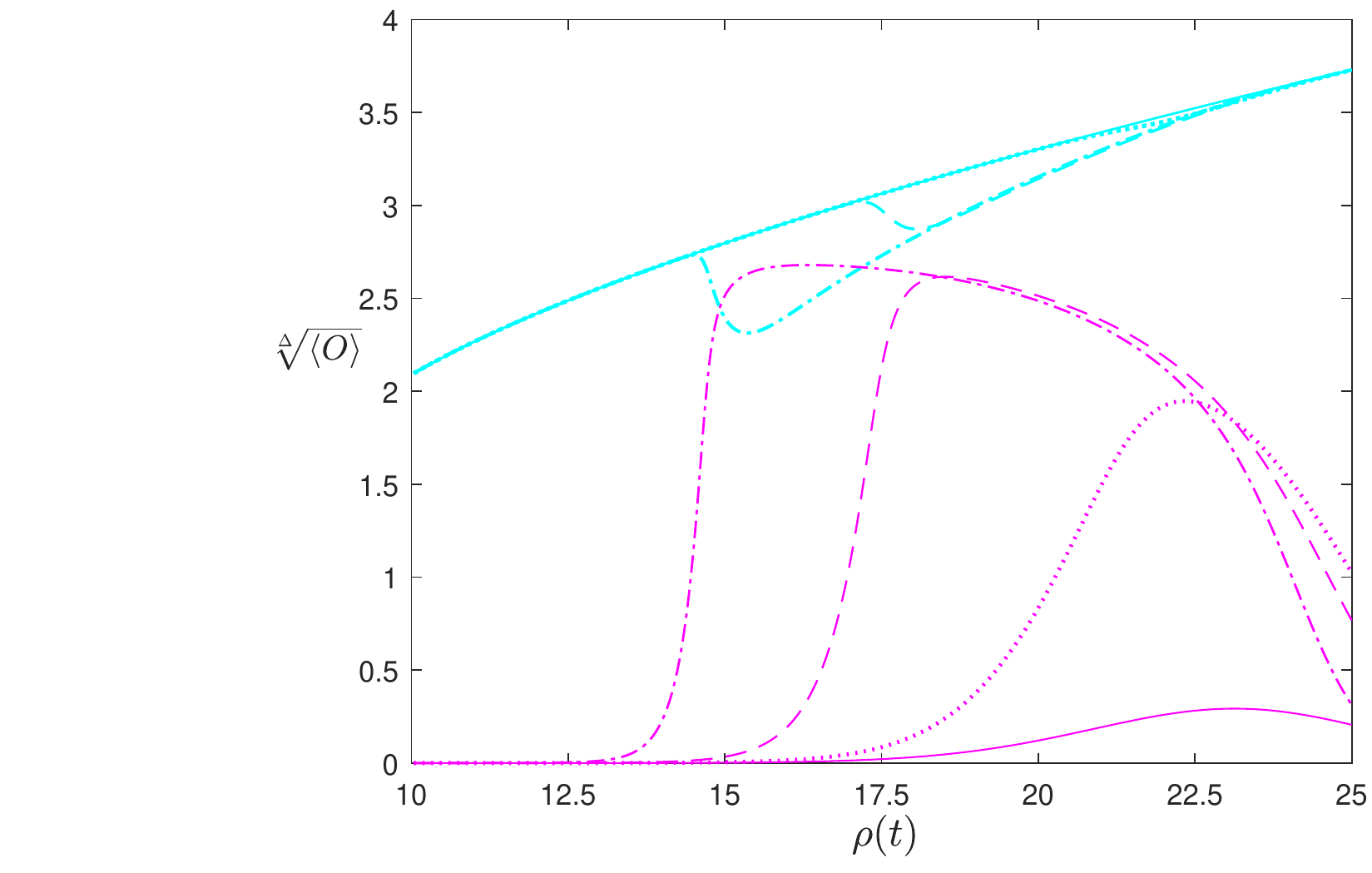}
\caption{\label{quench2} The values of condensate for s-wave and p-wave orders in quenching processes with different quench rate $v$. The horizontal axis show both the time coordinate $t$ and the value of $\rho(t)=t$, while the vertical axis show the condensate values $\sqrt[\Delta]{\langle O\rangle}$. Cyan represents S-wave order while magenta represents p-wave order. $q_p$ is fixed to $0.7189$. The dotted dashed lines, dashed lines and solid lines correspond to the processes with quench rates $v=0.00005$, $0.0001$ and $0.0002$, respectively.}
\end{figure}

To study the influence of quench rate $v$ on this kind of quenching processes, we fix $q_p=0.7189$ and plot the time dependent values of the s-wave and p-wave condensates in Figure.~\ref{quench2} for three values of quench rate $v$. In this case, we still set the horizontal axis as $\rho(t)$, 
therefore the different quenching processes are clearly presented in one plot, but the related time scale in this plot is different for the processes with different quench rate $v$.

In Figure.~\ref{quench2}, cyan represents for the condensate value of the s-wave order while magenta represents that for the p-wave order. The dotted dashed lines, dashed lines, dotted lines and solid lines correspond to the processes with quench rates $v=0.00005$, $0.0001$, $0.00015$ and $0.0002$, respectively. We can see that in the processes with different quench rate, the time dependent profiles of p-wave condensate value are different. In the case with a fast quench rate, the p-wave condensate can not increase to a large value in time; when the quench rate is a little slower, the p-wave order have enough time to grow to a large condensate value, but the decay from a large value also take more time; when the quench rate is very slow or even quasi-static, it will be almost along the condensate profile of static solution. Therefore, the dependence of the remaining p-wave condensate at the end of the quenching process on the quench rate $v$ is not monotonic.

\begin{figure}[tbp]
\includegraphics[width=0.8\textwidth,origin=c,trim=0 260 120 280,clip]{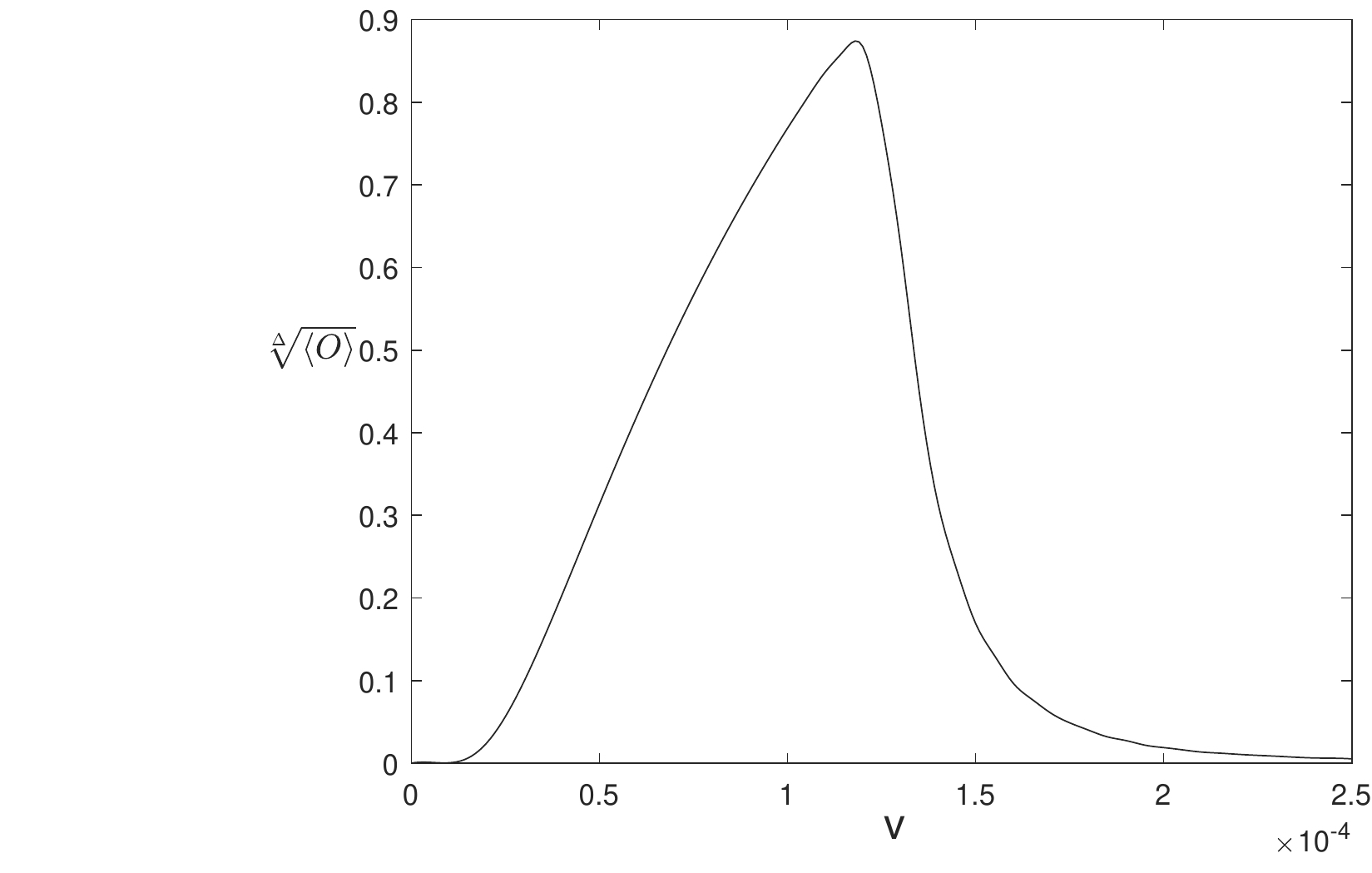}
\caption{\label{Opv} The remaining p-wave condensate $\sqrt[\Delta]{\langle O_{p_r}\rangle}$ at the end of the quenching process at different values of quench rate $v$ in quenching processes across the reentrant region.
}
\end{figure}

We show the dependence of the remaining p-wave condensate $\sqrt[\Delta]{\langle O_{p_r}\rangle}$ at the end of the quenching process $t=t_e$ on quench rate $v$ in Figure.~\ref{Opv}. We can see that as we expected, the remaining value of condensate for the p-wave order at the end of quenching process $t=t_e$ is non-monotonic and get a maximal value of $0.8741$ at $v=v_c=1.18\times 10^{-4}$.

\subsection{The evolution of competing orders in quenching from the normal phase}
In the previous quenching processes, the system start from the s-wave phase and the p-wave perturbation do not increase until the left critical point for the s+p phase. If we start from the normal state instead, the p-wave perturbation could increase before $\rho$ reaches the s+p region. In such a process, we can study the dynamical competition between the two different orders, and compare the evolution of the two condensates to that in the models with single order as well.


We set $q_p=0.7189$ and quench the system from a normal solution at $\rho_0=7.58$ to a final state with $\rho_f=7.65$, which is in the region dominated by the s-wave phase in the static phase diagram. Because the critical value $\rho_{cp}$ for the single condensate p-wave solution is a little larger but very close to the critical point $\rho_{cs}$ for the single condensate s-wave phase, the system go through the two critical points $\rho_{cs}$ and $\rho_{cp}$ successively, therefore the s-wave and p-wave condensates increase almost simultaneously. However, the p-wave condensate will finally decrease as a result of the final s-wave state. From these quenching processes, we get non-monotonic behavior of the p-wave condensate and the dynamical competition between the two orders.

\begin{figure}[tbp]
\includegraphics[width=0.47\textwidth,origin=c,trim=60 260 120 230,clip]{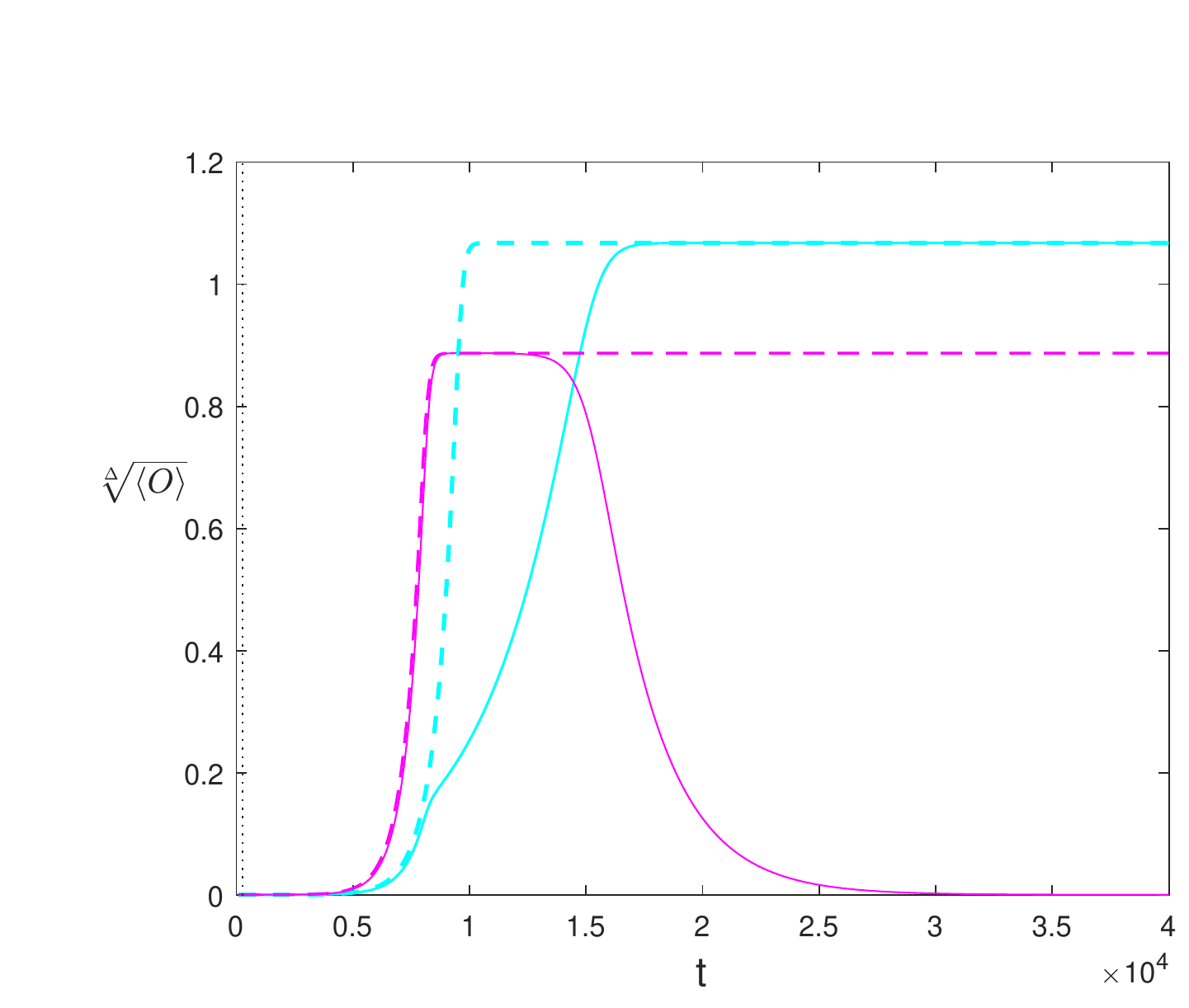}
\includegraphics[width=0.47\textwidth,origin=c,trim=60 260 120 230,clip]{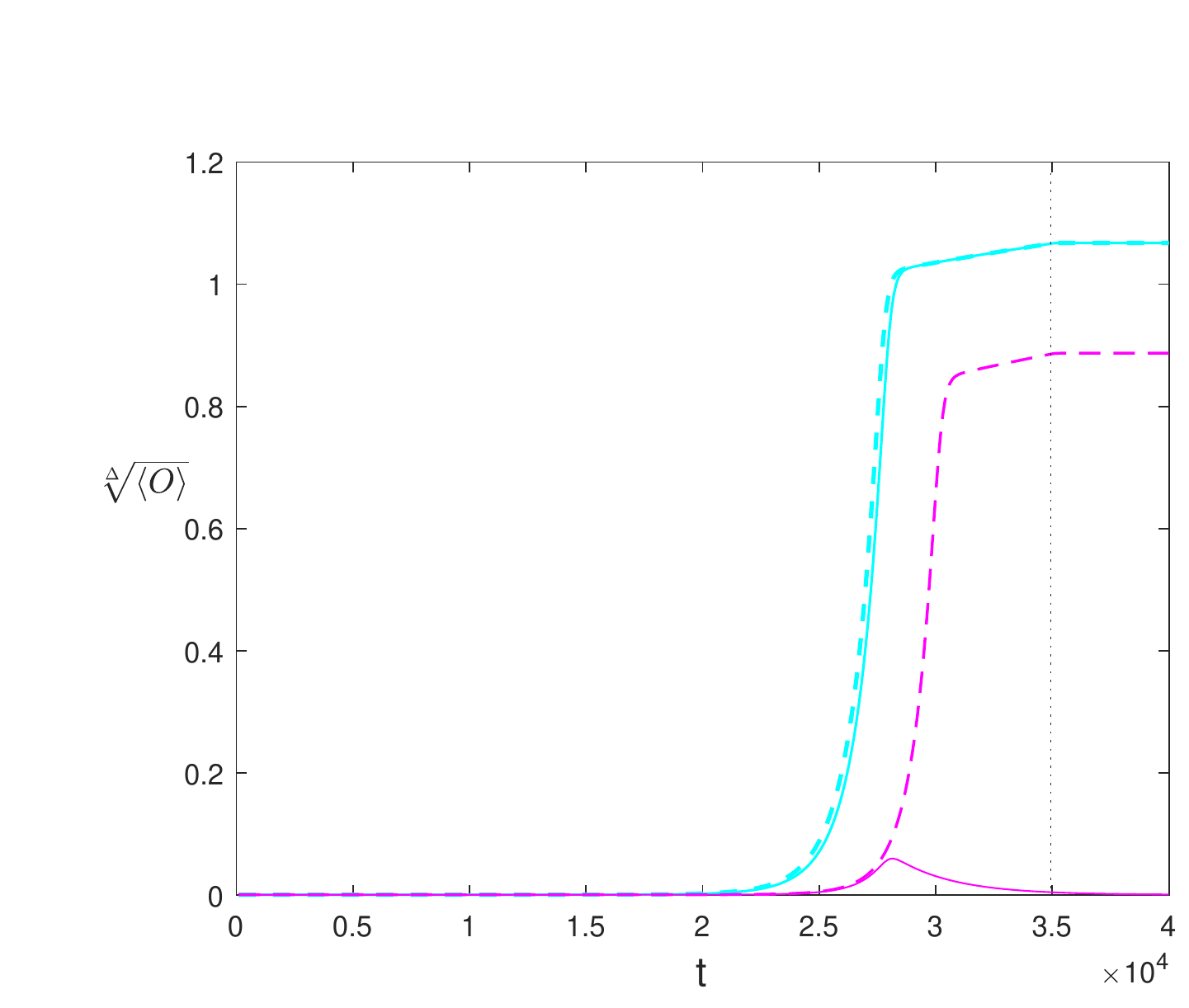}
\caption{\label{NtoS2} The condensate values in quenching processes from the normal phase with $q_p=0.7189$. The horizontal axis is time coordinate $t$, while the vertical axis is the condensate value $\sqrt[\Delta]{\langle O_{p_r}\rangle}$. Cyan represents s-wave order and magenta represents p-wave order. The solid lines show the evolution of the two condensate in the s+p model, while dashed lines show the evolution of the condensate in models with single order as comparison. The left plot is for $v=0.0002$ and the right plot is for $v=0.000002$. The dotted vertical lines denotes the end time of quenching $t_f$.}
\end{figure}

We show the time dependent condensate values in such quenching processes with two values of quench rate $v$ in Figure.~\ref{NtoS2}, where cyan represents s-wave condensate and magenta represents p-wave condensate. The solid lines show the evolution of the two condensate in the s+p model, while dashed lines show the evolution of the condensates in models with single order as comparison. 
The left plot is for $v=0.0002$ and the right plot is for $v=0.000002$, where the end time of quenching processes are $t_f=0.035\times 10^4$ and $t_f=3.5\times 10^4$, respectively. We can see from the two figures that in both cases, the p-wave condensate show a non-monotonic profile.

In the left plot in Figure.~\ref{NtoS2}, the quench rate is $v=0.0002$, and the quenching process end in a very shot time $t_f=0.035\times 10^4$. In this case the condensate value for the p-wave order grows even faster than the condensate value of s-wave order at beginning. Although the condensate values for the two different orders are different quantities, we can still claim that the p-wave order increase faster in the sense that the p-wave condensate get the maximum value earlier. The maximum value for the two orders are the condensate value for the single condensate solutions at $\rho=\rho_f=7.65$.

The evolution of the system can be roughly divided into three stages. The first stage is before the p-wave order reaches the maximum, in which the solid and dashed lines almost coincide. This can be explained by that the two orders do not coupled to each other directly, therefore before any of the two orders reaches the condensate value of single order static solution, the two orders growing as similar as in the cases with only single order turned on. 
In the second stage, after the p-wave order get the maximum condensate and before it begin to decrease, the s-wave order still increase because at $\rho=\rho_f$, the single order s-wave solution is the most stable one, but the increasing rate is slowed down, indicating the repelling effect between the two orders. In the final stage, the p-wave condensate is repelled by the increasing s-wave condensate and decrease from the maximum to zero.

In the right plot in Figure.~\ref{NtoS2}, we show a much slower quenching process with quench rate $v=0.000002$, which end at $t=t_f=3.5 \times 10^4$. In such a slow quenching process, we also divide the evolution into three stages.
The first stage is still before the p-wave order reaching the maximum, when the solid and dashed lines still almost coincide. However, we should notice that the dashed line for the s-wave condensate grows faster instead, therefore the solid lines show that in the first stage, the s-wave condensate grows faster as well.

Because the quench rate is slow enough, the two dashed lines reach the condensate value for static solution with $\rho=\rho(t)$ before the quenching stopped. Therefore the two dashed lines have two turning points. When the solid cyan line reaches the first turning point of the dashed cyan line, the p-wave condensate is suppressed and begin to decrease. This region with a decreasing p-wave condensate before the end of quenching $t_f$ is the second stage. The third stage is last section with $t>t_f$, where the system goes into equilibrium. Again because the quench rate is slow enough, the state at $t=t_f$ is very close to the static solution, and the s-wave condensate stopped increasing immediately.

We can conclude that: in both the fast and slow quenching processes from the normal phase, the two condensates grow freely in an early stage; after one condensate reaches the value for static solution, the growing of the other one is suppressed; finally, the p-wave condensate decrease to zero because the value $\rho$ of the final state is in the region dominated by the s-wave phase. As a result, the p-wave condensates show non-monotonic behavior with time again.

\begin{figure}[tbp]
\includegraphics[width=0.8\textwidth,origin=c,trim=0 260 120 280,clip]{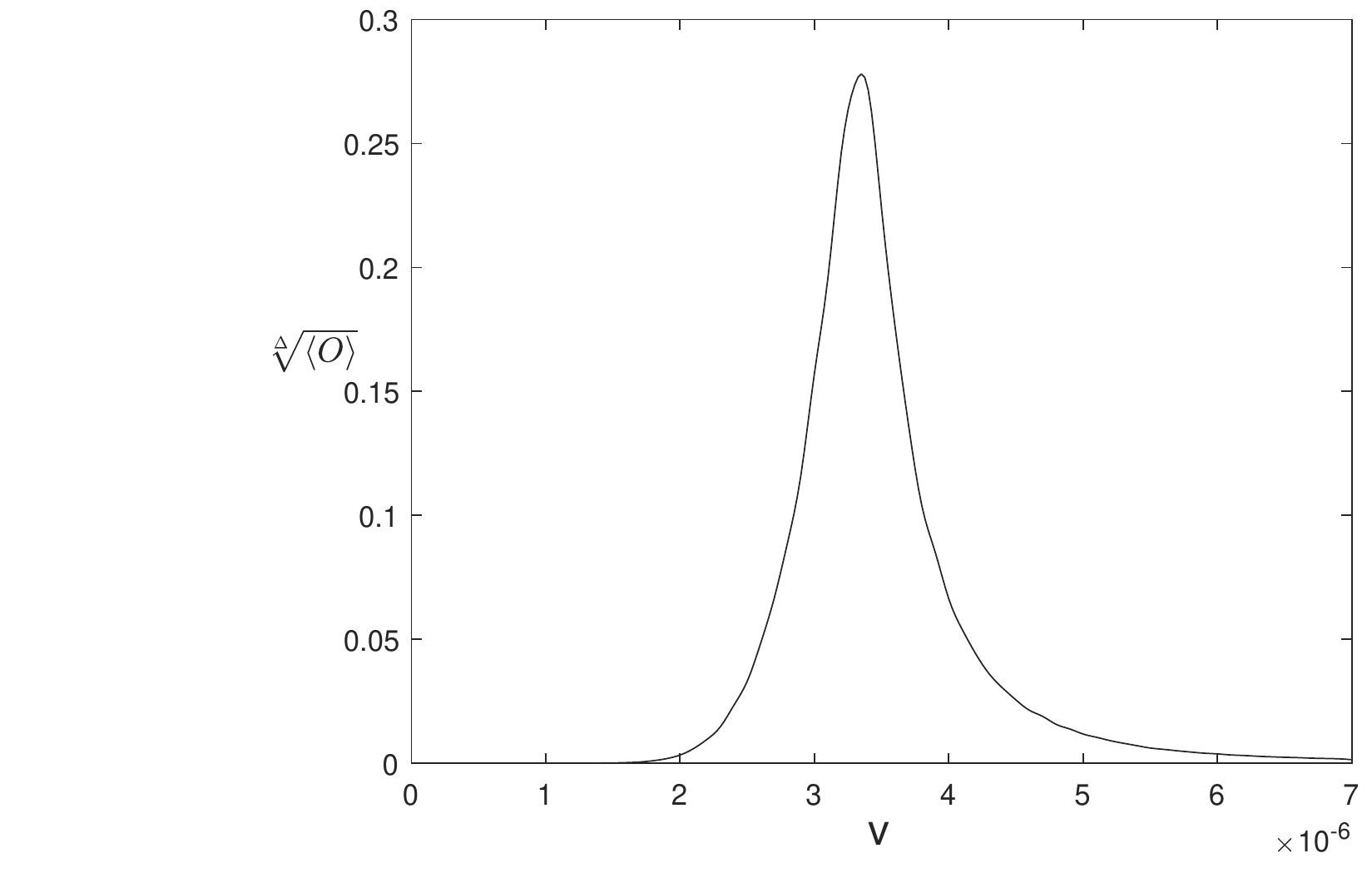}
\caption{\label{Opv2} The dependence of the remaining p-wave condensate $\sqrt[\Delta]{\langle O_{p_r}\rangle}$ at the end of the quenching process on the quench rate $v$ in quenching processes from the normal phase.
}
\end{figure}

We can see that the remaining p-wave condensate at the end of quenching $t=t_f$ in both the fast and slow quenching processes are very small. It is small in the fast quenching case because there is not enough time for the condensates to grow up in the very shot quenching period, while it is small in the slow case because the p-wave condensate already decrease to nearly zero in a long time. We again expect non-monotonic dependence of the remaining p-wave condensate on the quench rate and we draw this relation in Figure.~\ref{Opv2}.

In Figure.~\ref{Opv2}, we can see that the relation is indeed non-monotonic and the remaining p-wave condensate get a maximum value of $0.2781$ at $v=v_c=3.35\times 10^{-6}$.


\section{Conclusion and discussion}
\label{sect:conclusion}
In this paper, we study the holographic superconductivity model with an s-wave order and a p-wave order. We realize the reentrant phase transition without back-reaction and show a $q_p-\rho$ phase diagram. Based on these static phase structure, we take quenching processes to probe the dynamical properties across the reentrant region. We also study the quenching processes from normal phase to study the dynamical competition between the two orders and compare the evolution of two condensates to that in the model with single order.

In quenching processes across the reentrant region, we fix the initial and final states and quench the system with a linear time dependent function of charge density. Therefore these dynamical processes are tuned by the quench rate $v$ and $q_p$. We first fix the quench rate $v$ and compare quenching processes with different values of $q_p$ which controls the width of the reentrant region.  We see that the p-wave condensates show non-monotonic evolution, and the remaining value at the end of the quenching processes is larger with a larger value of $q_p$. Because the relaxing time near the final static state is also a increasing function on $q_p$, it takes more time to equilibrate with a larger value of $q_p$ after the end of quenching.

When $q_p$ is fixed and $v$ is varing, the different quenching processes take different time $t_f-t_0$ and the remaining p-wave condensate at the end of quenching $t=t_f$ is not monotonic on $v$.

In quenching processes from the normal phase, in an early stage before either of the two orders reaches the condensate value of static solution, we find that both the growth of s-wave and p-wave condensates are almost the same to the growth of condensates in models with single order. After one condensate reaches the condensate value of static solution, the growing of the other condensate is suppressed, indicating the repelling effect between the two condensates. Finally, because we stop quenching in the region dominated by the s-wave phase, the p-wave condensate decrease to zero and show a non-monotonic behavior with time.
We also draw the dependence of the remaining p-wave condensate at $t=t_f$ on the quench rate $v$, and show non-monotonic dependence which is similar to the relation in the quenching processes across the reentrant region.


The homogenous quenching processes studied in this paper already show rich phenomenon that need further investigation, such as the competition between the two condensates in dynamical processes and non-monotonic dependence of the remaining p-wave condensate on the quench rate $v$, which is discovered in two different sets of quenching processes. We are going to further study these phenomenon and potential critical behaviors in future.

\acknowledgments
ZYN would like to thank Yu Tian for helpful discussions.
This work is partially supported by the National Natural Science Foundation of China under Grant Nos. 11965013, 11675140 and 11565017.
YZ is partially supported by the Fund for Reserve Talents of Young and Middle-aged Academic and Technical Leaders of Yunnan Province (Grant No. 2018HB006). ZYN and YZ are partially supported by Yunnan Ten Thousand Talents Plan Young \& Elite Talents Project.




\end{document}